%% file: draft.tex
\documentclass[ALICE,manyauthors]{cernphprep}
\usepackage[comma,square,numbers,sort&compress]{natbib}
\usepackage{hyperref}
\usepackage[utf8]{inputenc}
\usepackage[T1]{fontenc}
\usepackage{hyperref}
\usepackage{lineno}
\usepackage{color}
\usepackage{xspace}
\usepackage{orcidlink}

\def\PbPb  {\mbox{Pb--Pb}\xspace}
\def\OO  {\mbox{OO}\xspace}
\def\NeNe  {\mbox{Ne--Ne}\xspace}
\def\XeXe  {\mbox{Xe--Xe}\xspace}
\def\pPb   {\mbox{p--Pb}\xspace}

\def\pp    {\mbox{pp}\xspace}

\newcommand{\pt}{\ensuremath{p_{\rm T}}\xspace}

\newcommand{\Ne}{\ensuremath{^{20}\rm Ne}\xspace}
\newcommand{\Oxy}{\ensuremath{^{16}\rm O}\xspace}

\graphicspath{{.img/}}

\usepackage{verbatim}

\begin{document}%
\begin{titlepage}
\PHyear{2025}       
\PHnumber{203}      
\PHdate{02 Sep}  

\title{Evidence of nuclear geometry-driven anisotropic flow in \OO and \NeNe collisions at $\mathbf{\sqrt{{\textit s}_{\rm\mathbf {NN}}}}$ = 5.36 TeV}
\ShortTitle{Anisotropic flow in \OO and \NeNe}   

\Collaboration{ALICE Collaboration\thanks{See Appendix~\ref{app:collab} for the list of collaboration members}}
\ShortAuthor{ALICE Collaboration} 

\begin{abstract}
A central question in strong-interaction physics, governed by quantum chromodynamics (QCD), is whether femto-scale droplets of quark--gluon plasma (QGP) form in small collision systems involving projectiles significantly smaller than heavy ions. Collisions of light ions such as $^{16}$O and $^{20}$Ne offer a unique opportunity to probe the emergence of collective behavior in QCD matter.
This Letter presents the first measurements of elliptic ($v_2$) and triangular ($v_3$) flow of charged particles in $^{16}$O--$^{16}$O and $^{20}$Ne--$^{20}$Ne collisions at a center-of-mass energy per nucleon pair of $\sqrt{s_{_{\rm NN}}} = 5.36$ TeV with the ALICE detector.
The hydrodynamic model predictions, explicitly incorporating the nuclear structures of $^{16}$O and $^{20}$Ne, exhibit a good agreement with the flow measurements presented. The observed increase of $v_2$ in central \NeNe collisions relative to \OO collisions, driven by the nuclear geometries, highlights the importance of utilizing light nuclei with well-defined geometric shapes to constrain the initial conditions. These findings support the presence of nuclear geometry-driven hydrodynamic flow in light-ion collisions at the LHC.
\end{abstract}

\end{titlepage}
                                                                                                                                     
\setcounter{page}{2}  

The primary goal of ultrarelativistic nuclear collisions is to recreate the quark--gluon plasma (QGP) --- a state of matter that is thought to have existed in the early Universe --- and to investigate its properties~\cite{STAR:2005gfr,PHENIX:2004vcz,PHOBOS:2004zne,ALICE:2022wpn,CMS:2024krd}. A significant effort has been dedicated to determining the precise properties and time evolution of the QGP. One of the most important approaches is the study of anisotropic flow~\cite{Ollitrault:1992bk}, which arises from the transfer of initial spatial anisotropy to momentum anisotropy in the final state via pressure gradients within the created medium. Anisotropic flow is characterized by the Fourier coefficients $v_n$ of the azimuthal particle distribution,
\begin{equation}
    \frac{dN}{d\varphi} \propto 1+2\sum\limits_{n=1}\limits^{\infty}{v_n\cos[n(\varphi-\Psi_n)]},
    \label{eq:FourierSeries}
\end{equation}
where $\varphi$ is the azimuthal angle of the final state produced particle, $\Psi_n$ is the $n^{\rm th}$-order symmetry plane, and $v_n$ are the flow coefficients.
Experimental measurements of $v_{n}$ coefficients and their event-by-event fluctuations, from two- and multi-particle correlations with inclusive and identified charged hadrons in heavy-ion collisions, have been performed~\cite{ALICE:2011ab,ALICE:2014wao,ALICE:2016ccg, ALICE:2018lao, ALICE:2019zfl, ALICE:2022zks, ATLAS:2012at,ATLAS:2019dct, CMS:2013wjq,CMS:2019cyz}. 
The measurements can be quantitatively reproduced by hydrodynamic model calculations. This agreement provides strong evidence that the QGP created in heavy-ion collisions at the LHC behaves as a nearly perfect fluid~\cite{Heinz:2013th, Bernhard:2016tnd,JETSCAPE:2020shq, Nijs:2020ors}, characterized by an exceptionally small specific shear viscosity ($\eta/s$), which approaches the quantum lower bound of $1/4\pi$~\cite{Kovtun:2004de}.

In addition to colliding heavy ions at ultrarelativistic energies, the LHC also provides pp and p--Pb collisions. The formation of a QGP was not anticipated in these small systems and they were initially viewed as baselines for heavy-ion collisions~\cite{Citron:2018lsq}.
Surprisingly, the experimental results  have revealed new manifestations of emergent QCD phenomena, such as finite anisotropic flow in high-multiplicity events~\cite{CMS:2010ifv, ALICE:2013snk, CMS:2014und, ATLAS:2015hzw, PHENIX:2018lia, CMS:2018loe, ALICE:2019zfl, ALICE:2022wpn, ALICE:2023gyf, ALICE:2025bwp}, with the first indication provided by the observation of a long-range ‘ridge’ in two-particle correlations in pp and p–Pb collisions at the LHC~\cite{CMS:2010ifv, ALICE:2013snk, CMS:2014und, ATLAS:2015hzw}, in p--Au at RHIC~\cite{STAR:2015kak,PHENIX:2014fnc} and even possible hints in high multiplicity $e^+e^-$ collisions at ALEPH~\cite{Electron-PositronAlliance:2023klx}. Subsequent multiparticle correlation measurements confirmed the collective nature of the signal~\cite{PHENIX:2018lia, CMS:2018loe, ALICE:2019zfl}. Measurements performed in different collision systems, such as p--Au, d--Au, and $^{3}$He--Au at RHIC~\cite{PHENIX:2018lia, STAR:2022pfn}, as well as pp, p--Pb, and Pb--Pb collisions at the LHC~\cite{ALICE:2019zfl,ALICE:2023gyf,ALICE:2024vzv, ALICE:2025bwp, CMS:2015yux, CMS:2016fnw, ATLAS:2017rtr}, revealed that the observed anisotropic flow in small collision systems is predominantly driven by the initial geometry and its event-by-event fluctuations through the complex dynamic evolution of the created system~\cite{Song:2017wtw,Nagle:2018nvi,Christiansen:2024bhe,Grosse-Oetringhaus:2024bwr}.
The most recent anisotropic flow measurements for identified hadrons presented results consistent with the formation of partonic collectivity in small systems~\cite{ALICE:2024vzv,Zhao:2020wcd}.

While hydrodynamic and parton transport models qualitatively describe many of the collective flow features in small collision systems, there are discrepancies, i.e., in the magnitudes of flow coefficients, between model calculations and experimental measurements~\cite{Bozek:2011if, Mantysaari:2017cni, Weller:2017tsr, Zhao:2017rgg, Zhao:2020pty, He:2015hfa, Kurkela:2018ygx}. 
These differences largely arise from the limited understanding of the initial conditions in pp and p–Pb collisions, in particular, the spatial and subnucleonic structure of the proton~\cite{YuanyuanWang:2023wbz,Zhao:2020wcd,Schenke:2020mbo,Mantysaari:2022ffw}. 
In contrast, in light-ion collisions, such as \OO and \NeNe, the flow is predicted to be predominantly driven by the shape (eccentricity) of the initial overlap region~\cite{Zhang:2017xda,Giacalone:2024luz,Li:2020vrg,YuanyuanWang:2024sgp,Zhao:2024feh,Lim:2018huo,Summerfield:2021oex,Behera:2023nwj}. These geometric features are more precisely characterized when the colliding nuclei have been systematically investigated~\cite{Jia:2022ozr,Giacalone:2025vxa}.
Even though the ground states of \Oxy and \Ne have $J^\pi=0^+$, and are therefore spherically symmetric in the laboratory frame, their wave functions can be described as superpositions of intrinsic shapes averaged over all orientations~\cite{Ring1980}. 
In modern nuclear-structure models, such intrinsic shapes emerge as a tetrahedral 4-$\alpha$\footnote{$\alpha$ particle is a tightly bound state of two protons and two neutrons} configuration for $^{16}$O, while $^{20}$Ne resembles an $\alpha+^{16}$O ``bowling pin''--like structure~\cite{Lee:2025req,Frosini:2021fjf,Frosini:2021sxj} as a consequence of the spatial many-body nucleon correlations. As the even-even $J^\pi=0^+$ nuclei are rotationally invariant in the laboratory frame, any mapping of measured flow to ground-state deformations remains inherently model-dependent and should be considered complementary to traditional low-energy nuclear-structure constraints~\cite{Dobaczewski:2025rdi}.

As an example, central heavy-ion collisions were proven to be particularly sensitive to the shape of the nucleus, such as the quadrupole deformation and triaxiality of $^{129}$Xe and $^{192}$Au nuclei~\cite{STAR:2015mki,STAR:2024wgy,ALICE:2021gxt,ATLAS:2022dov,ALICE:2024nqd}. 
Given that \OO and \NeNe have similar system sizes~\cite{Giacalone:2024luz}, the evolution of the two systems is similar, so that the ratios between the two systems substantially mitigate final-state effects (e.g. hadronic rescattering, jet, decay) and thus highlight the differences in the effects of the initial conditions~\cite{Giacalone:2024luz,Jia:2022ozr,Giacalone:2025vxa,Zhang:2020hww}, i.e. nuclear shape, subnucleonic structure, and initial fluctuations.

In this Letter, the anisotropic flow measurements are performed utilizing data from the short light-ion runs, OO and Ne--Ne collisions, delivered by the LHC in July 2025 and recorded with the ALICE apparatus~\cite{ALICE:2023udb}. Similar measurements have been performed at ATLAS around the same time~\cite{ATLAS:2025nnt} and shortly after by the CMS Collaboration~\cite{CMS:2025tga}. 
The centrality of the collisions, which quantifies the fraction of a geometrical cross section of the colliding nuclei, is estimated using the multiplicity measured in the Fast Interaction Trigger detector, located at a pseudorapidity range of $-3.3<\eta<-2.1$ (FT0C)~\cite{ALICE:2023udb}. 
Events are required to be within 10 cm of the nominal interaction point.
To reject pileup events and optimize the centrality estimation, bunch crossings with more than one reconstructed vertex are rejected.
In total, about 3 billion \OO collisions and 400 million \NeNe collisions pass these criteria. 

The charged particles are reconstructed based on information from the Inner Tracking System (ITS)~\cite{ALICE:2013nwm, ALICE:2023udb} and the Time Projection Chamber (TPC)~\cite{ALICETPC:2020ann}. High-quality tracks are selected by requiring more than 50 associated TPC clusters, at least 70 crossed rows in the TPC, and 5 associated ITS clusters, as well as a $\chi^2$ per degree of freedom of the track fit to the TPC space points less than 2.5. Contamination from secondary particles is largely reduced by requiring tracks to have a distance of closest approach (DCA) to the primary vertex of less than 2 cm along the longitudinal direction and a transverse DCA less than a $\pt$-dependent value ranging from 0.2 cm at $\pt=0.2$ GeV/\textit{c} to 0.02 cm at $\pt=3$ GeV/\textit{c}. 

The flow measurements presented in this Letter were obtained from two- and multiparticle cumulants using the generic framework~\cite{Bilandzic:2013kga,Moravcova:2020wnf}, which corrects for detector acceptance and track reconstruction efficiency effects. To suppress nonflow contamination, which are the azimuthal angle correlations not associated with the $n^{\rm th}$-order harmonic symmetry plane angle $\Psi_{\rm n}$, a pseudorapidity separation of $|\Delta\eta|>1.4$ is applied in the two-particle correlation measurements~\cite{Zhou:2015iba}. 
The four-particle correlations are, by construction less sensitive to nonflow effects~\cite{Moravcova:2020wnf}. Thus, the standard method is applied by default, while the subevent method~\cite{Huo:2017nms} was applied as a cross check.

Systematic uncertainties are estimated with relative differences by varying the event and track selection criteria, and are added to the estimations of the remaining nonflow effects. Uncertainties related to the event selection include the variation in the accepted vertex position along the beam line.
The resulting systematic uncertainty was found to be within 2\%.
In the tracking procedure, the variations of the maximum allowed DCA to the primary vertex, both along the beam line and in the transverse plane, result in differences of less than 1\%. 
To estimate the remaining nonflow contamination, the dihadron correlation method~\cite{CMS:2010ifv,ALICE:2012eyl} with $|\Delta\eta| > 1.4$, together with an additional template fit approach, was examined~\cite{ATLAS:2015hzw,ALICE:2024vzv}. The deviations from the cumulant measurements are treated as systematic uncertainties associated with nonflow effects.
Only the sources of systematic uncertainties found to be statistically significant by more than 1$\sigma$ following the procedure in Ref.~\cite{Barlow:2002yb} are combined in quadrature to obtain the total systematic uncertainty. 
As a result, systematic uncertainties up to 6.5\% are assigned to $v_n\{2\}$ and $v_2\{4\}$ for both \OO and \NeNe collisions, and systematic uncertainties within 1.5\% are assigned to the ratio $v_n\{2\}$(\NeNe/\OO) and $v_2\{4\}$(\NeNe/\OO).

\begin{figure}[htb!]
    \centering
    \includegraphics[width=0.6\linewidth]{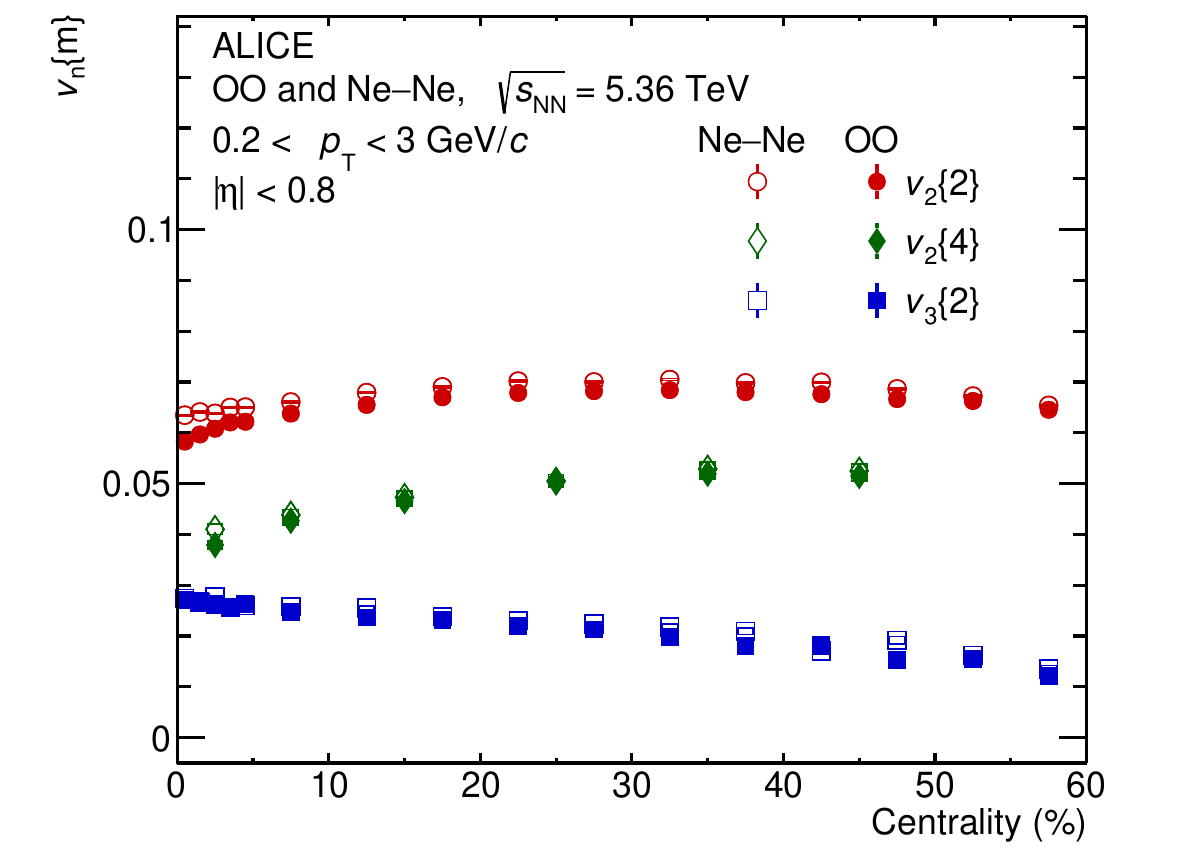}
    \caption{Charged particle $v_2\{2\}$, $v_3\{2\}$, and $v_2\{4\}$ as a function of centrality in \OO (solid markers) and \NeNe (open markers) collisions at $\sqrt{s_{_\mathrm{NN}}}$ = 5.36 TeV. The vertical lines represent statistical uncertainties, and the open boxes represent systematic uncertainties, with most of them being smaller than the symbol size.}
    \label{fig:vnData}
\end{figure}
The centrality dependence of $v_2\{2\}$, $v_3\{2\}$, and $v_2\{4\}$ in \OO and \NeNe collisions at $\sqrt{s_{_\mathrm{NN}}} = 5.36$ TeV is shown in Fig.~\ref{fig:vnData}.
In both collision systems, $v_{2}\{2\}$ exhibits a weak centrality dependence; it initially increases with increasing centrality percentile up to approximately 30\%, following the trend of the initial state eccentricity $\varepsilon_2$~\cite{Nijs:2021clz}. Then, a decrease of the measured $v_{2}\{2\}$ in more peripheral collisions is observed.
The $v_{2}\{4\}$ measurement also shows an increasing trend with increasing centrality percentile for the presented centrality region, chosen as the region where the non-flow of $v_2\{4\}$ is under control, and the $v_2\{4\}$ is expected to decrease at higher centralities similar to $v_2\{2\}$. Smaller values of $v_2\{4\}$ compared to $v_{2}\{2\}$, and the nonzero value of the measured $v_{2}\{4\}$ indicates the collective nature of the measured anisotropy and confirms the presence of anisotropic flow in \OO and \NeNe collisions. Assuming that the applied pseudorapidity gap effectively suppresses nonflow effects in two-particle correlations, the remaining differences between two- and four-particle cumulants of $v_{2}$ can be attributed to event-by-event fluctuations of elliptic flow, which are further quantified in Fig.~\ref{fig:v24overv22} in the Supplemental Material.
The observed increasing trend of $v_{2}\{4\}$ with increasing centrality percentile signifies the system’s response to the changes of ``lenticular'' shape formed by the two overlapping nuclei. An increasing trend of $v_{2}\{4\}$ toward more peripheral collisions (lower multiplicity) was not previously observed in either pp or in p--Pb collisions~\cite{CMS:2013jlh,ALICE:2019zfl}, but is observed in Fig.~\ref{fig:vnData} for both \OO and \NeNe collisions reflecting the geometric initial conditions of these collisions.

In contrast to the $v_{2}$ results, $v_{3}\{2\}$ decreases with larger centrality percentiles.
This decreasing trend and the magnitudes of $v_{3}\{2\}$ are similar to what is seen in the $v_{3}\{2\}$ measurements in \pp and \pPb collisions~\cite{ALICE:2019zfl}.
This is expected as $v_{3}\{2\}$ arises from initial state fluctuations, which are comparable at the same multiplicity in small systems~\cite{CMS:2016fnw,ALICE:2019zfl}. This is also consistent with low-multiplicity measurements of $v_3$ in \XeXe and \PbPb collisions~\cite{Aaboud:2017blb}, confirming the fluctuation-driven nature of $v_3$ in non-central collisions across system sizes.

\begin{figure}[htb!]
    \centering
    \includegraphics[width=0.8\linewidth]{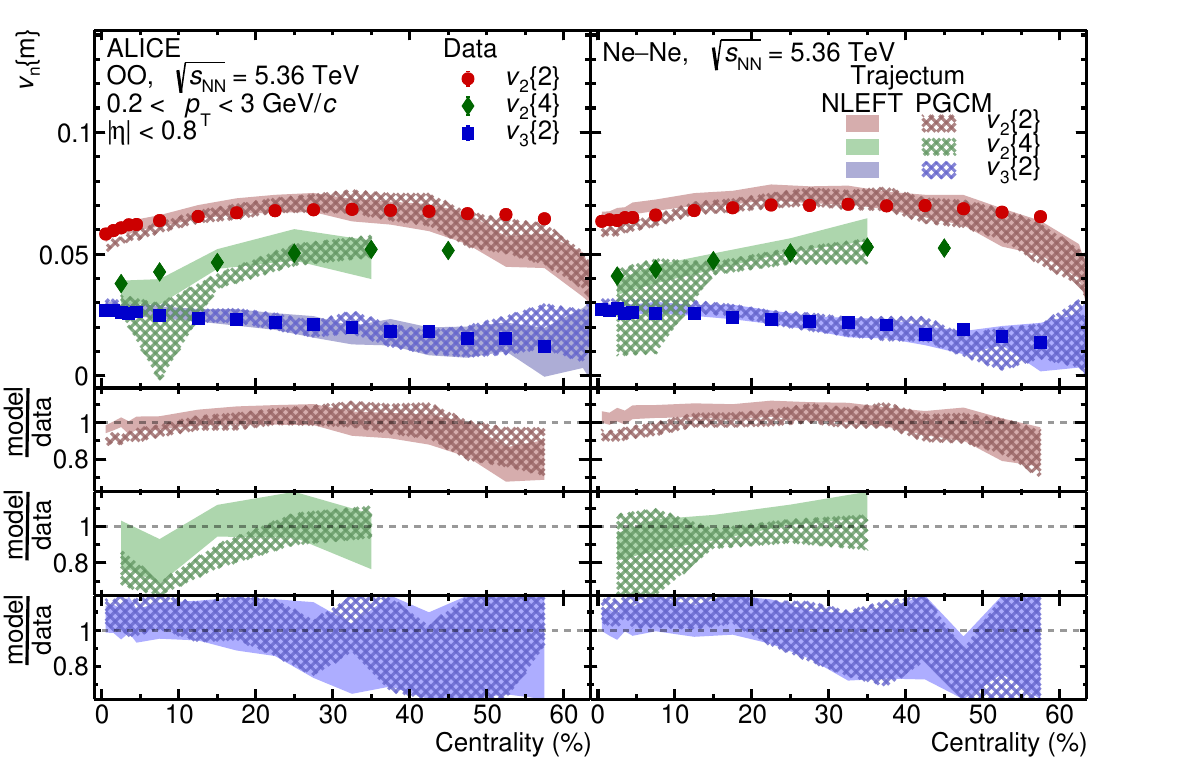}
    \caption{Charged particle $v_2\{2\}$, $v_3\{2\}$, and $v_2\{4\}$ as a function of centrality in \OO (left) and \NeNe (right) collisions at $\sqrt{s_{_\mathrm{NN}}}$ = 5.36 TeV. The vertical lines represent statistical uncertainties, and the open boxes represent systematic uncertainties, with most of them being smaller than the symbol size. The measurements are compared with Trajectum calculations with NLEFT and PGCM inputs~\cite{Giacalone:2024luz}.}
    \label{fig:vnNeNeOO}
\end{figure}

In Fig.~\ref{fig:vnNeNeOO}, the measurements are compared with hydrodynamic simulations from the Trajectum framework~\cite{Giacalone:2024luz, Nijs:2021clz, Nijs:2020roc}, which incorporate nuclear structure inputs for $^{16}$O and $^{20}$Ne derived from the Nuclear Lattice Effective Field Theory (NLEFT)~\cite{Lee:2025req} and the {\it ab initio} Projected Generator Coordinate Method (PGCM)~\cite{Frosini:2021fjf,Frosini:2021sxj}. Both approaches model ground-state nucleon correlations within the nuclei and represent the current best understanding of the nuclear structure. The NLEFT approach constructs nuclear structures from first principles by evolving nucleon configurations via Monte Carlo simulations on a discrete lattice, naturally incorporating many-body correlations and emergent clustering. The PGCM instead starts from energy-minimized mean-field states and includes collective correlations by projecting onto appropriate quantum numbers and mixing configurations across deformation spaces.
In both frameworks, as mentioned previously, the \Ne nuclei exhibit a bowling‑pin-like $\Oxy+\alpha$ structure, while the \Oxy nuclei have irregular tetrahedron-like shapes~\cite{Giacalone:2024luz}. 
The theoretical predictions are shown as colored curves, with bands indicating the combined statistical and systematic uncertainties. 
The goal of these comparisons is to assess whether the combination of realistic nuclear structures and a hydrodynamic framework tuned using measurements from heavy-ion collisions can quantitatively predict the experimental measurements in light-ion collisions.
As shown in Fig.~\ref{fig:vnNeNeOO}, NLEFT-based calculations successfully reproduce both the trend and magnitude of $v_{2}\{2\}$, $v_{3}\{2\}$, and $v_{2}\{4\}$ up to 50\% centrality. The agreements with respect to the flow measurements equal or surpass the accuracy of full Bayesian parameter extraction in previous heavy-ion studies, which have have been based on one-body nuclear density distribution of the structure of heavy nuclei (i.e. $^{129}$Xe, $^{208}$Pb) ~\cite{Bernhard:2016tnd,JETSCAPE:2020shq,Nijs:2020ors}. The agreement between the experimental measurements and theoretical calculations presented in this letter suggests that developing realistic low-energy ab initio calculations for heavier nuclei could significantly improve our understanding of the initial conditions of nuclear collisions, and therefore, enhance the precision of the extraction of QGP properties. In the case of PGCM-based calculations, the agreement is slightly worse in central collisions, though the deviation from the flow measurements remains within a few percent. In noncentral collisions, the calculations are consistent with the measurements within the uncertainties. In particular, the $v_{3}\{2\}$ results are highly nontrivial, as the measurements clearly exhibit a different centrality dependence from the initial-state triangular eccentricity $\varepsilon_{3}$ calculation, which predicts larger values in semicentral than in central collisions~\cite{Nijs:2021clz}. 
This difference between the initial-state geometric shape and the final-state triangular flow highlights the role of the strong response of the created system, which is accurately captured by the hydrodynamic predictions. 
It is noteworthy that when centrality is estimated using particle multiplicity in the midrapidity region (which is used in the Trajectum calculations), the $v_{2}\{2\}$ and $v_{3}\{2\}$ measurements remain unchanged, while a $\sim$10\% reduction in $v_{2}\{4\}$ is observed, leading to a better agreement with hydrodynamic predictions.  The measured $v_2\{4\}$ is accurately described by the hydrodynamic predictions in both \OO and \NeNe. This is the first time the measurements of multi-particle cumulant of anisotropic flow can be successfully reproduced by hydrodynamic models in small systems. Notably, the long-standing sign-problem present in hydrodynamic descriptions of pp collisions~\cite{Zhao:2017rgg,Zhao:2020pty} is not observed in light-ion collisions.
The consistency between the presented $v_{2}\{2\}$, $v_{3}\{2\}$, and $v_{2}\{4\}$ measurements and hydrodynamic model predictions supports the emergence of collective behavior in light-ion collisions, well described by a hydrodynamic expansion of the QGP.

\begin{figure}[!htb]
    \centering
    \includegraphics[width=0.5\textwidth]{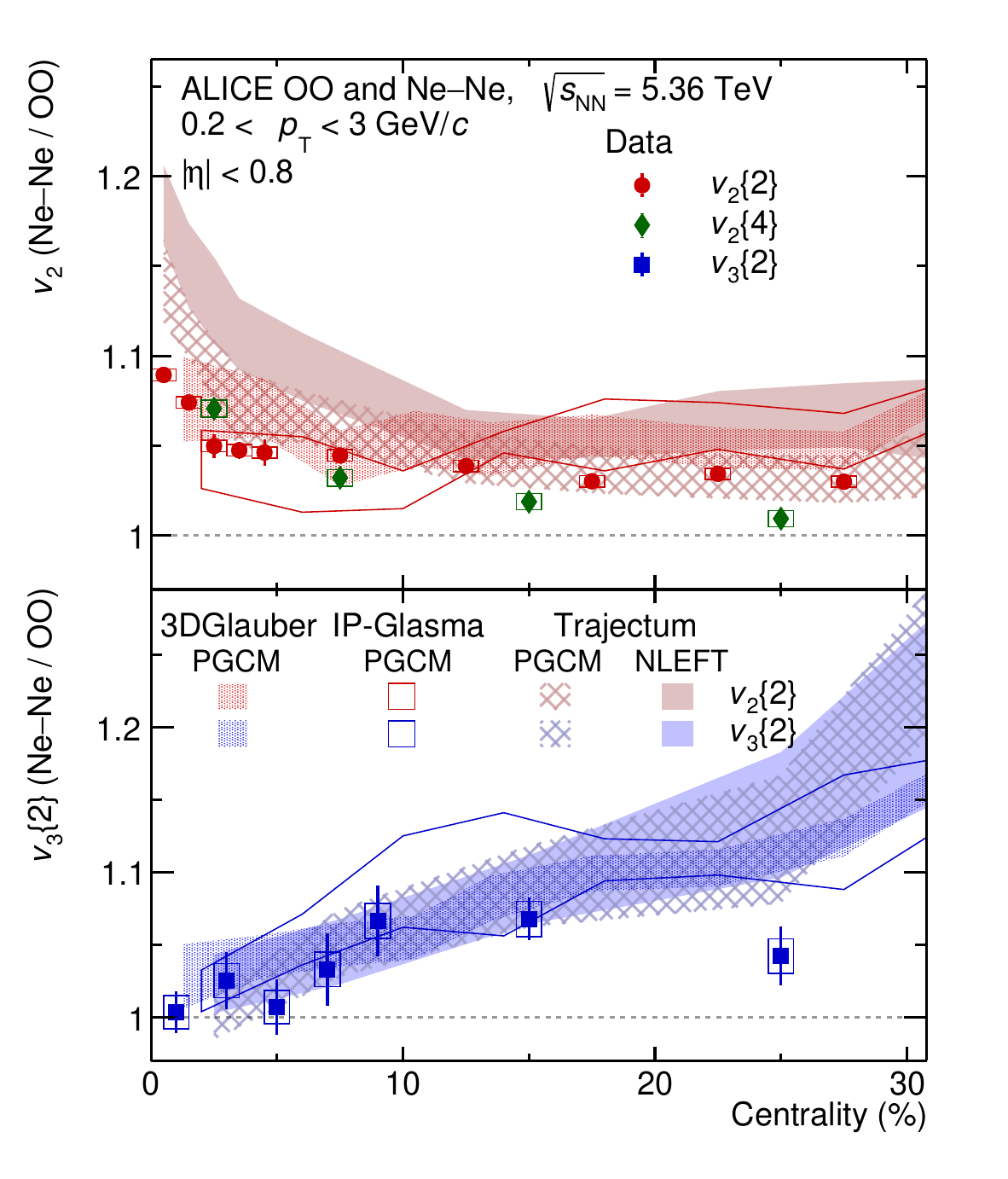}
  \caption[.]{Ratios $v_2$(Ne--Ne/OO) and $v_3$(Ne--Ne/OO) as a function of centrality in the 0–30\% centrality range. The vertical lines represent statistical uncertainties and the open boxes represent the systematic uncertainties, while most of them are smaller than the symbol size. The measurements are compared with Trajectum calculations with NLEFT and PGCM inputs~\cite{Giacalone:2024luz} as well as IP-Glasma+JIMWLK+MUSIC+UrQMD~\cite{Mantysaari:2025tcg} and 3DGlauber+MUSIC+UrQMD~\cite{Jahan:2024wpj} calculations with PGCM input.}
    \label{fig:vnratio}
\end{figure}
Ratios between two collision systems of comparable size are particularly valuable, as they largely suppress final-state effects and reduce systematic uncertainties for both theoretical model calculations and experimental measurements~\cite{Giacalone:2021uhj,STAR:2021mii}. These ratios serve as sensitive probes of the initial-state geometry and enable rigorous testing of the theoretical understanding of the dynamic evolution in small systems. The centrality dependence of the system-comparison ratios $v_{2}\{2\}$(\NeNe/\OO) and $v_{3}\{2\}$(\NeNe/\OO) is presented in Fig.~\ref{fig:vnratio}. 
For $v_{2}\{2\}$(\NeNe/\OO), the ratio peaks at around 1.08 in ultracentral collisions and rapidly decreases to about 1.05 by 10\% centrality, after which it exhibits a weak centrality dependence. Similar centrality dependence is also observed in the ratio of $v_{2}\{4\}$(\NeNe/\OO). This pronounced enhancement in central events is attributed to the stronger quadrupole deformation of $^{20}$Ne compared to the tetrahedral $^{16}$O~\cite{Giacalone:2024luz}. 
In contrast, $v_{3}\{2\}$(\NeNe/\OO) begins around unity in central collisions and increases gradually with increasing centrality percentile, reaching approximately 1.06 at 10\%. The smaller $v_{3}\{2\}$ ratio in central collisions could be attributed to the tetrahedral configuration of $^{16}$O, which leads to a larger octupole deformation. This results in a larger triangular eccentricity $\varepsilon_3$ and consequently, a smaller $v_{3}\{2\}$(\NeNe/\OO) in the most central collisions.

As shown in Fig.~\ref{fig:vnNeNeOO}, the measurements of the centrality dependence of anisotropic flow favor the NLEFT-based hydrodynamic calculations. In contrast, the system-comparison ratios are closer to the PGCM-based predictions.
Notably, both NLEFT- and PGCM-based hydrodynamic calculations slightly overestimate the measured $v_2\{2\}$ ratios as seen in Fig.~\ref{fig:vnratio}.
Given the fact that the two systems evolve similarly, leading to a suppression of many final-state effects when taking the ratio, this major discrepancy may originate from the modeling of the immediate post-collision initial state. Specifically, parameters governing energy deposition, nucleon and subnucleon spatial profiles, and preequilibrium dynamics may require refinement~\cite{Giacalone:2025vxa}. It is not known a priori whether gluon distributions and saturation in ultra‑relativistic collisions affect nuclear structure. Since such effects are not considered in low‑energy descriptions but may become important at high energies, the measurements provide a lever for exploring potential differences in nuclear structure across energy scales. 

To determine whether the observed discrepancies in the system ratio stem from the initial conditions employed in hydrodynamic calculations or from inaccurate nuclear structure inputs, calculations from both the IP-Glasma+JIMWLK+MUSIC+UrQMD model~\cite{Mantysaari:2025tcg} and the 3DGlauber+MUSIC+UrQMD framework~\cite{Jahan:2024wpj} are presented. Both models utilize nuclear structure inputs from {\it ab initio} PGCM~\cite{Frosini:2021fjf,Frosini:2021sxj}, identical to those used in the Trajectum predictions.
As shown in Fig.~\ref{fig:vnratio}, the IP-Glasma-based hydrodynamic model describes the system ratios very well, capturing both the trend and the magnitude of the measurements.
A comparison to the individual flow coefficients can be found in Fig.~\ref{fig:AMPT}.
The main difference in the system ratio calculations between the Trajectum and IP-Glasma+JIMWLK+MUSIC+UrQMD models most likely lies in their description of the initial conditions immediately after the collisions and early-time dynamics prior to thermalization. 
The subnucleon width $w_q$, the effective transverse size over which partons (quarks and gluons) inside a nucleon are distributed, is obtained from Bayesian inference from \PbPb collisions and has a size of $w_q \approx 0.40$ fm for the Trajectum model. It is much larger compared to the IP-Glasma+JIMWLK+MUSIC+UrQMD model, which uses a width of $w_q\approx 0.11$ fm inferred from incoherent $J/\psi$ production~\cite{Mantysaari:2016ykx,Mantysaari:2016jaz} from HERA data~\cite{Caldwell:2010zza,Schenke:2012hg}.  
At the same time, a reasonable agreement is also seen between the measurements and the 3DGlauber+MUSIC+UrQMD calculations~\cite{Jahan:2024wpj}, which use a $w_q \approx 0.11$ fm.
The determination of the nucleon and subnucleon width with the Bayesian parameter estimation framework has so far proved to be challenging~\cite{Giacalone:2021clp}, and only recently has the nucleon width been better constrained in Bayesian models~\cite{Mantysaari:2025ltq}. The present measurements favor the calculations with a smaller subnucleon width of approximately 0.1--0.2 fm, as previously suggested~\cite{Giacalone:2025vxa}. Although other processes or assumptions of the initial-state models may contribute to the differences, such measurements provide additional constraints for improving the description of the early-time structure and dynamics of the QGP, and the inclusion of the ratio measurements in future Bayesian analyses will allow for a more precise extraction of the sub-nucleon width of the light nuclei.

This Letter presents the first measurements of elliptic flow ($v_2$) and triangular flow ($v_3$) in \OO and \NeNe collisions at $\sqrt{s_{_{\rm NN}}} = 5.36$ TeV using the ALICE detector. 
The flow coefficients exhibit characteristic centrality dependence in both collision systems, which was not observed previously in pp and p--Pb collisions. 
The combination of nuclear structure inputs from NLEFT and PGCM model together with the standard modeling framework of relativistic heavy-ion collisions, the hydrodynamic model, shows a ‌reasonable agreement with the anisotropic flow measurements in a wide range of centralities. Importantly, the NLEFT and PGCM models utilise the many-body nuclear densities to determine the distribution of nucleons~\cite{Giacalone:2024luz} as emphasised in Ref.~\cite{Dobaczewski:2025rdi}. 
The agreement between theoretical predictions and experimental measurements reported in this Letter reveals the emergence of hydrodynamic flow across collision systems at the LHC, effectively bridging the gap between heavy-ion collisions and proton--proton and proton--nucleus collisions. 
Moreover, the ratios between \OO and \NeNe collisions substantially reduce the influence from the final-state effects and serve as precise probes of the initial conditions. The pronounced enhancement of $v_{2}\{2\}$(\NeNe/\OO) observed in central collisions is attributed to the significant nuclear geometric effects of $^{16}$O and $^{20}$Ne nuclei on the final-state flow coefficients. The comparison of the measured $v_{2}\{2\}$(\NeNe/\OO) with the various hydrodynamic model calculations, including state of the art nuclear structure, points to imperfect configurations of the initial conditions, particularly the sub-nucleon size.
These precision measurements, therefore, offer a unique probe into the early-time dynamics and (sub)structure of the colliding nuclei. They open new avenues for imaging the nuclear structure of light ions in the near future, complementary to low-energy nuclear-structure experiments such as the $\alpha$-cluster configurations in $^{16}$O and $^{20}$Ne, which have profound implications for understanding quantum many-body phenomena, a longstanding challenge in nuclear physics. 

\section*{Data availability}

This manuscript has associated data in a HEPData repository at Ref.~\cite{hepdata.179652},

\newenvironment{acknowledgement}{\relax}{\relax}
\begin{acknowledgement}
\section*{Acknowledgments}
The ALICE Collaboration would like to thank Giuliano Giacalone, Govert Nijs, Wilke van der Schee (Trajectum), Chun Shen, Bjoern Schenke, Wenbin Zhao (IP-Glasma+MUSIC+UrQMD),  Chun Shen, Wenbin Zhao (3D-Glauber+ MUSIC+UrQMD), Ziwei Lin, Guoliang Ma, Xinli Zhao (AMPT) for providing the latest predictions.

\input{fa_2025-09-01_Opt_C.tex}
\end{acknowledgement}

\bibliographystyle{utphys}
\bibliography{bibliography}
\newpage
\appendix

\section{Supplemental Material}
\label{app:SM}
This section presents the ratios of $v_3\{2\}/v_2\{2\}$ and $v_2\{4\}/v_2\{2\}$, together with the double ratios comparing both collision systems.

\begin{figure}[!htb]
    \begin{center}
      \includegraphics[width=0.5\textwidth]{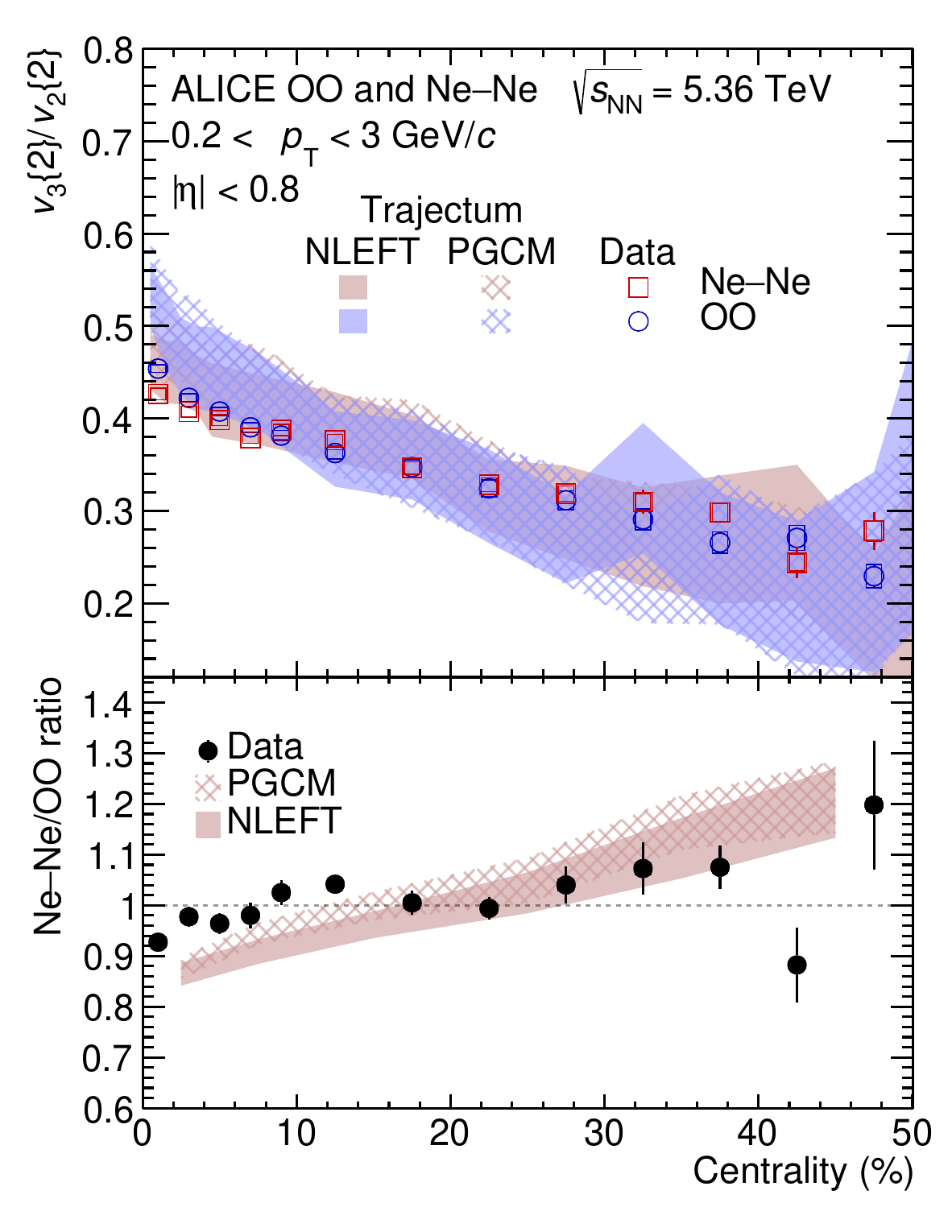}
      \caption[.]{(top): Ratio of the charged particle $v_3\{2\}$ to $v_2\{2\}$ as a function of centrality in OO and Ne--Ne collisions at $\sqrt{s_{_\mathrm{NN}}}$ = 5.36 TeV.
      (bottom): Double ratios $v_3\{2\} / v_2\{2\}$(Ne--Ne/OO). The vertical lines represent statistical uncertainties and the open boxes represent the systematic uncertainties while most of them are smaller than the symbol size. The measurements are compared with Trajectum calculations with NLEFT and PGCM inputs~\cite{Giacalone:2024luz}.}
      \label{fig:v32overv22}
    \end{center}
\end{figure}
The predicted shapes for $^{16}$O and $^{20}$Ne nuclei have different quadrupole and octupole deformations. In other words, $^{16}$O has stronger octupole deformation because of its tetrahedral shape, whereas $^{20}$Ne has stronger quadrupole deformation because of its ``bowling pin "-like shape.
Therefore, the ratio of $v_3\{2\} / v_2\{2\}$ in \OO collisions is expected to be larger than that in \NeNe collisions. As shown in Fig.~\ref{fig:v32overv22}, this expectation is confirmed only in the most central collisions, where the nuclear geometric effect is strongest. Beyond this region, the ratio of $v_3\{2\} / v_2\{2\}$ shown in the top panel exhibits a decreasing trend with increasing centrality in both collision systems, with the ratio in \OO collisions remaining slightly higher than in \NeNe collisions in the most central collisions.
Both NLEFT- and PGCM-based hydrodynamic calculations describe the measurements, demonstrating the impact of the unique nuclear shape imprinted by the $\alpha$-cluster structures in \Oxy and \Ne. 

\begin{figure}[!htb]
    \begin{center}
      \includegraphics[width=0.5\textwidth]{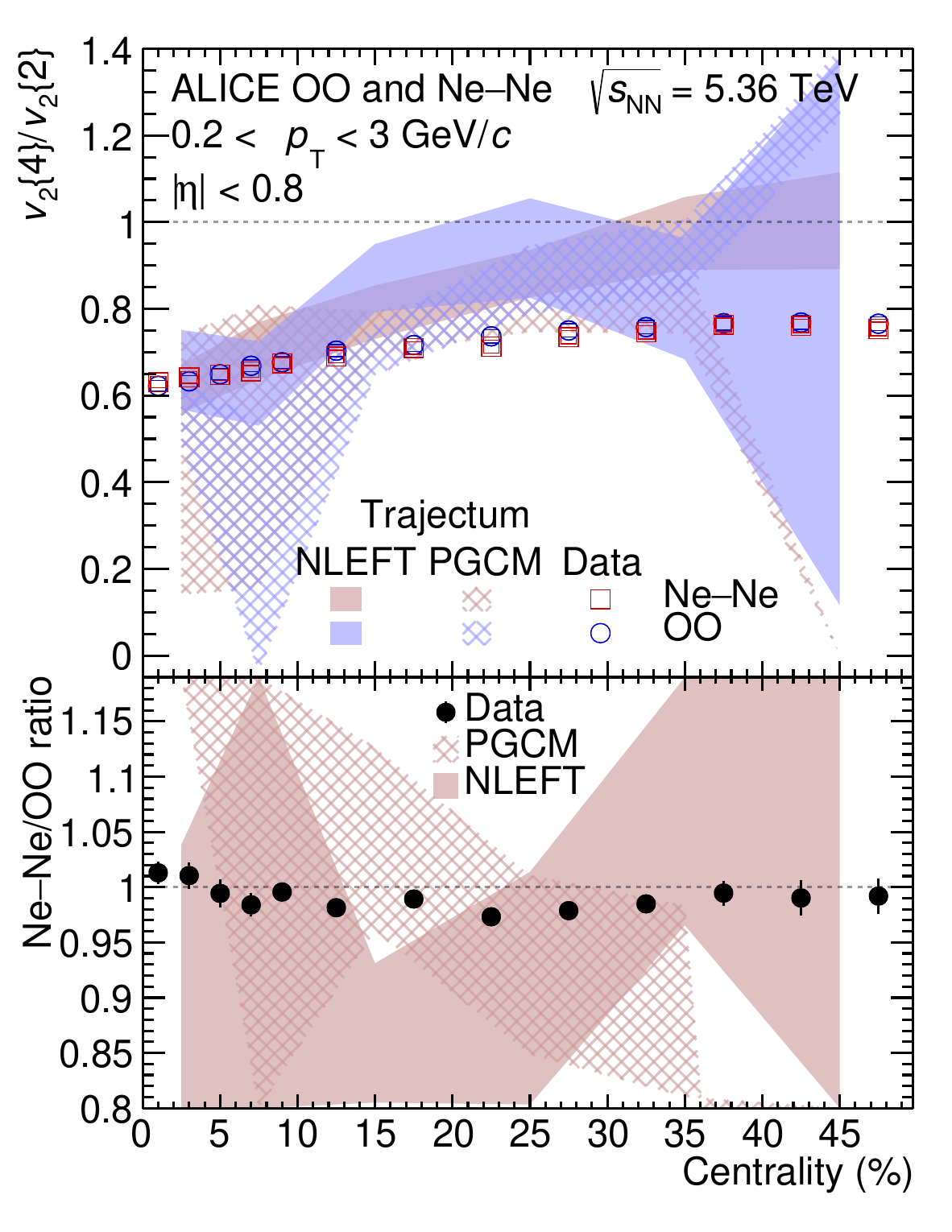}
      \caption[.]{(top): Ratio of the charged particle $v_2\{4\}$ to $v_2\{2\}$ as a function of centrality in OO and Ne--Ne collisions at $\sqrt{s_{_\mathrm{NN}}}$ = 5.36 TeV.
      (bottom): Double ratios $v_2\{4\} / v_2\{2\}$(Ne--Ne/OO). The vertical lines represent statistical uncertainties and the open boxes represent the systematic uncertainties while most of them are smaller than symbol size. The measurements are compared to Trajectum calculations with NLEFT and PGCM inputs~\cite{Giacalone:2024luz}.}
      \label{fig:v24overv22}
    \end{center}
\end{figure}
In the hydrodynamic evolution picture, elliptic flow arises from the anisotropic expansion of the QGP, which is driven by the spatial distribution of nucleons in the colliding nuclei. 
The $v_{2}$ from two- and four-particle cumulants carries different sensitivities to the event-by-event flow fluctuations, originating from the initial geometry fluctuations. 
The ratio of $v_2\{4\}$/$v_2\{2\}$ minimizes the effects of the system evolution and highlights the initial fluctuations. Previous studies have shown that the ratio is sensitive to the $\alpha$-cluster configuration and compactness of \Oxy in relativistic \OO collisions~\cite{Chen:2024xbi,YuanyuanWang:2024sgp,Zhao:2024feh,Zhang:2024vkh}.
In Fig.~\ref{fig:v24overv22}, the centrality dependence of $v_2\{4\}$/$v_2 \{2\}$ in \OO and \NeNe collisions, as well as the ratio between the two collision systems, are presented. The $v_2\{4\}$/$v_2\{2\}$ is lowest in central collisions and increases with larger centrality. A similar trend is observed in both \XeXe~\cite{ALICE:2018lao} and \PbPb collisions~\cite{ALICE:2016ccg}, albeit with larger magnitudes as the larger flow fluctuations in the small light-ion collisions reduce the $v_2\{4\}$/$v_2\{2\}$ ratio. The double ratio of $v_2\{4\}$/$v_2 \{2\}$ between \NeNe and \OO is slightly below unity in noncentral collisions due to the larger flow fluctuations in the smaller \OO collisions. In central collisions, the double ratio increases due to the effect of nuclear structure. The \Ne nuclei have larger deformation and thus \NeNe collisions produce stronger $v_2$ than \OO collisions, while the flow fluctuations remain similar. Note that the $v_2\{4\}/v_2\{2\}$ ratio in \OO and \NeNe would decrease by a few percent if centrality is redefined from an estimation based on the multiplicity at forward rapidity (default option) to one based on midrapidity, consistent with what is discussed in Fig.~\ref{fig:vnNeNeOO}.

\begin{figure}[!htb]
    \begin{center}
      \includegraphics[width=0.8\textwidth]{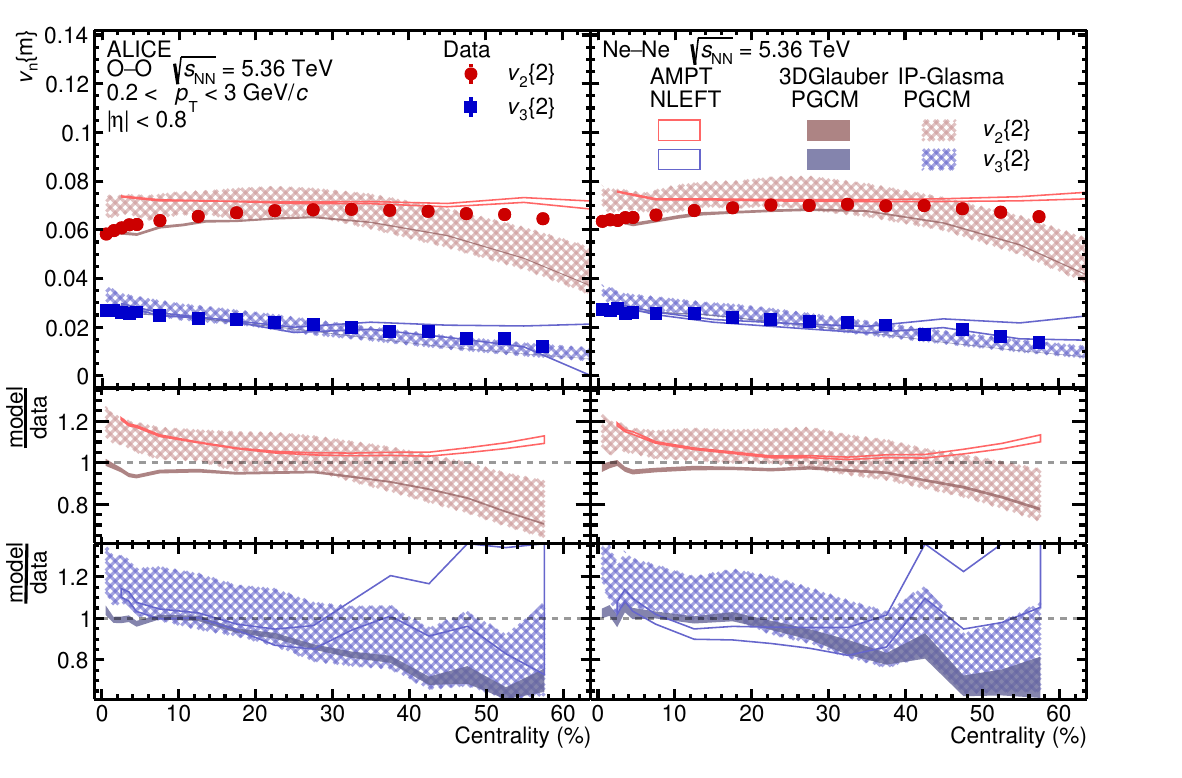}
      \caption[.]{Charged particle $v_2\{2\}$ and $v_3\{2\}$ as a function of centrality in \OO (left) and \NeNe (right) collisions. The data points are the same as those in Fig.~\ref{fig:vnNeNeOO}. The measurements are compared with IP-Glasma+JIMWLK+MUSIC+UrQMD~\cite{Mantysaari:2025tcg} and 3DGlauber+MUSIC+UrQMD~\cite{Jahan:2024wpj} calculations with PGCM input, as well as AMPT~\cite{Zhao:2024feh} with NLEFT input.}
      \label{fig:AMPT}
    \end{center}
\end{figure}

In Fig.~\ref{fig:AMPT}, the measurements of flow coefficients, identical to those shown in Fig.~\ref{fig:vnNeNeOO}, are compared with hydrodynamic calculations from the 3D-Glauber+MUSIC+UrQMD and IP-Glasma+JIMWLK+MUSIC +UrQMD frameworks, as well as with calculations from the AMPT transport model. The hydrodynamic calculations are both based on PGCM nuclear structure inputs, whereas the AMPT model uses nuclear structure from NLEFT. The IP-Glasma-based calculations overestimate the $v_{2}$ measurements in the most central collisions 0–10\%, underestimate them at centralities above 50\%, and show reasonable agreement with the data across intermediate centralities. In contrast, the 3D-Glauber-based calculations consistently underestimate the $v_{2}$ measurements, whereas the AMPT model shows no clear centrality dependence and systematically overestimates the $v_{2}$ values across the entire centrality range. All presented model calculations exhibit reasonable agreement with the measured $v_{3}$, although the uncertainties remain large.

\clearpage
\section{The ALICE Collaboration}
\label{app:collab}

\input{Alice_Authorlist_2025-09-01_Opt_C.tex}

\end{document}

%% file: fa_2025-09-01_Opt_C.tex

The ALICE Collaboration would like to thank all its engineers and technicians for their invaluable contributions to the construction of the experiment and the CERN accelerator teams for the outstanding performance of the LHC complex.
The ALICE Collaboration gratefully acknowledges the resources and support provided by all Grid centres and the Worldwide LHC Computing Grid (WLCG) collaboration.
The ALICE Collaboration acknowledges the following funding agencies for their support in building and running the ALICE detector:
A. I. Alikhanyan National Science Laboratory (Yerevan Physics Institute) Foundation (ANSL), State Committee of Science and World Federation of Scientists (WFS), Armenia;
Austrian Academy of Sciences, Austrian Science Fund (FWF): [M 2467-N36] and Nationalstiftung f\"{u}r Forschung, Technologie und Entwicklung, Austria;
Ministry of Communications and High Technologies, National Nuclear Research Center, Azerbaijan;
Rede Nacional de Física de Altas Energias (Renafae), Financiadora de Estudos e Projetos (Finep), Funda\c{c}\~{a}o de Amparo \`{a} Pesquisa do Estado de S\~{a}o Paulo (FAPESP) and The Sao Paulo Research Foundation  (FAPESP), Brazil;
Bulgarian Ministry of Education and Science, within the National Roadmap for Research Infrastructures 2020-2027 (object CERN), Bulgaria;
Ministry of Education of China (MOEC) , Ministry of Science \& Technology of China (MSTC) and National Natural Science Foundation of China (NSFC), China;
Ministry of Science and Education and Croatian Science Foundation, Croatia;
Centro de Aplicaciones Tecnol\'{o}gicas y Desarrollo Nuclear (CEADEN), Cubaenerg\'{\i}a, Cuba;
Ministry of Education, Youth and Sports of the Czech Republic, Czech Republic;
The Danish Council for Independent Research | Natural Sciences, the VILLUM FONDEN and Danish National Research Foundation (DNRF), Denmark;
Helsinki Institute of Physics (HIP), Finland;
Commissariat \`{a} l'Energie Atomique (CEA) and Institut National de Physique Nucl\'{e}aire et de Physique des Particules (IN2P3) and Centre National de la Recherche Scientifique (CNRS), France;
Bundesministerium f\"{u}r Forschung, Technologie und Raumfahrt (BMFTR) and GSI Helmholtzzentrum f\"{u}r Schwerionenforschung GmbH, Germany;
General Secretariat for Research and Technology, Ministry of Education, Research and Religions, Greece;
National Research, Development and Innovation Office, Hungary;
Department of Atomic Energy Government of India (DAE), Department of Science and Technology, Government of India (DST), University Grants Commission, Government of India (UGC) and Council of Scientific and Industrial Research (CSIR), India;
National Research and Innovation Agency - BRIN, Indonesia;
Istituto Nazionale di Fisica Nucleare (INFN), Italy;
Japanese Ministry of Education, Culture, Sports, Science and Technology (MEXT) and Japan Society for the Promotion of Science (JSPS) KAKENHI, Japan;
Consejo Nacional de Ciencia (CONACYT) y Tecnolog\'{i}a, through Fondo de Cooperaci\'{o}n Internacional en Ciencia y Tecnolog\'{i}a (FONCICYT) and Direcci\'{o}n General de Asuntos del Personal Academico (DGAPA), Mexico;
Nederlandse Organisatie voor Wetenschappelijk Onderzoek (NWO), Netherlands;
The Research Council of Norway, Norway;
Pontificia Universidad Cat\'{o}lica del Per\'{u}, Peru;
Ministry of Science and Higher Education, National Science Centre and WUT ID-UB, Poland;
Korea Institute of Science and Technology Information and National Research Foundation of Korea (NRF), Republic of Korea;
Ministry of Education and Scientific Research, Institute of Atomic Physics, Ministry of Research and Innovation and Institute of Atomic Physics and Universitatea Nationala de Stiinta si Tehnologie Politehnica Bucuresti, Romania;
Ministerstvo skolstva, vyskumu, vyvoja a mladeze SR, Slovakia;
National Research Foundation of South Africa, South Africa;
Swedish Research Council (VR) and Knut \& Alice Wallenberg Foundation (KAW), Sweden;
European Organization for Nuclear Research, Switzerland;
Suranaree University of Technology (SUT), National Science and Technology Development Agency (NSTDA) and National Science, Research and Innovation Fund (NSRF via PMU-B B05F650021), Thailand;
Turkish Energy, Nuclear and Mineral Research Agency (TENMAK), Turkey;
National Academy of  Sciences of Ukraine, Ukraine;
Science and Technology Facilities Council (STFC), United Kingdom;
National Science Foundation of the United States of America (NSF) and United States Department of Energy, Office of Nuclear Physics (DOE NP), United States of America.
In addition, individual groups or members have received support from:
Czech Science Foundation (grant no. 23-07499S), Czech Republic;
FORTE project, reg.\ no.\ CZ.02.01.01/00/22\_008/0004632, Czech Republic, co-funded by the European Union, Czech Republic;
European Research Council (grant no. 950692), European Union;
Deutsche Forschungs Gemeinschaft (DFG, German Research Foundation) ``Neutrinos and Dark Matter in Astro- and Particle Physics'' (grant no. SFB 1258), Germany;
CONVECS project, CUP C97H23001700002 FESR 2021-2027 program, Italy;
FAIR - Future Artificial Intelligence Research, funded by the NextGenerationEU program (Italy).

%% file: Alice_Authorlist_2025-09-01_Opt_C.tex
\begin{flushleft} 
\small

I.J.~Abualrob\,\orcidlink{0009-0005-3519-5631}\,$^{\rm 114}$, 
S.~Acharya\,\orcidlink{0000-0002-9213-5329}\,$^{\rm 50}$, 
G.~Aglieri Rinella\,\orcidlink{0000-0002-9611-3696}\,$^{\rm 32}$, 
L.~Aglietta\,\orcidlink{0009-0003-0763-6802}\,$^{\rm 24}$, 
N.~Agrawal\,\orcidlink{0000-0003-0348-9836}\,$^{\rm 25}$, 
Z.~Ahammed\,\orcidlink{0000-0001-5241-7412}\,$^{\rm 134}$, 
S.~Ahmad\,\orcidlink{0000-0003-0497-5705}\,$^{\rm 15}$, 
I.~Ahuja\,\orcidlink{0000-0002-4417-1392}\,$^{\rm 36}$, 
Z.~Akbar$^{\rm 81}$, 
A.~Akindinov\,\orcidlink{0000-0002-7388-3022}\,$^{\rm 140}$
% ...

\section*{Affiliation Notes}

$^{\rm I}$ Deceased\\
$^{\rm II}$ Also at: Max-Planck-Institut fur Physik, Munich, Germany\\
$^{\rm III}$ Also at: Czech Technical University in Prague, Prague, Czech Republic\\
$^{\rm IV}$ Also at: Instituto de Fisica da Universidade de Sao Paulo\\
$^{\rm V}$ Also at: Dipartimento DET del Politecnico di Torino, Turin, Italy\\
$^{\rm VI}$ Also at: University College of Dublin, Dublin, Ireland\\
$^{\rm VII}$ Also at: Department of Applied Physics, Aligarh Muslim University, Aligarh, India\\
$^{\rm VIII}$ Also at: Institute of Theoretical Physics, University of Wroclaw, Wroclaw, Poland\\
$^{\rm IX}$ Also at: Facultad de Ciencias, Universidad Nacional Aut\'{o}noma de M\'{e}xico, Mexico City, Mexico\\

\section*{Collaboration Institutes}

$^{1}$ A.I. Alikhanyan National Science Laboratory (Yerevan Physics Institute) Foundation, Yerevan, Armenia\\
$^{2}$ AGH University of Krakow, Cracow, Poland\\
$^{3}$ Bogolyubov Institute for Theoretical Physics, National Academy of Sciences of Ukraine, Kyiv, Ukraine\\
$^{4}$ Bose Institute, Department of Physics  and Centre for Astroparticle Physics and Space Science (CAPSS), Kolkata, India\\
$^{5}$ California Polytechnic State University, San Luis Obispo, California, United States\\
$^{6}$ Central China Normal University, Wuhan, China\\
$^{7}$ Centro de Aplicaciones Tecnol\'{o}gicas y Desarrollo Nuclear (CEADEN), Havana, Cuba\\
$^{8}$ Centro de Investigaci\'{o}n y de Estudios Avanzados (CINVESTAV), Mexico City and M\'{e}rida, Mexico\\
$^{9}$ Chicago State University, Chicago, Illinois, United States\\
$^{10}$ China Nuclear Data Center, China Institute of Atomic Energy, Beijing, China\\
$^{11}$ China University of Geosciences, Wuhan, China\\
$^{12}$ Chungbuk National University, Cheongju, Republic of Korea\\
$^{13}$ Comenius University Bratislava, Faculty of Mathematics, Physics and Informatics, Bratislava, Slovak Republic\\
$^{14}$ Creighton University, Omaha, Nebraska, United States\\
$^{15}$ Department of Physics, Aligarh Muslim University, Aligarh, India\\
$^{16}$ Department of Physics, Pusan National University, Pusan, Republic of Korea\\
$^{17}$ Department of Physics, Sejong University, Seoul, Republic of Korea\\
$^{18}$ Department of Physics, University of California, Berkeley, California, United States\\
$^{19}$ Department of Physics, University of Oslo, Oslo, Norway\\
$^{20}$ Department of Physics and Technology, University of Bergen, Bergen, Norway\\
$^{21}$ Dipartimento di Fisica, Universit\`{a} di Pavia, Pavia, Italy\\
$^{22}$ Dipartimento di Fisica dell'Universit\`{a} and Sezione INFN, Cagliari, Italy\\
$^{23}$ Dipartimento di Fisica dell'Universit\`{a} and Sezione INFN, Trieste, Italy\\
$^{24}$ Dipartimento di Fisica dell'Universit\`{a} and Sezione INFN, Turin, Italy\\
$^{25}$ Dipartimento di Fisica e Astronomia dell'Universit\`{a} and Sezione INFN, Bologna, Italy\\
$^{26}$ Dipartimento di Fisica e Astronomia dell'Universit\`{a} and Sezione INFN, Catania, Italy\\
$^{27}$ Dipartimento di Fisica e Astronomia dell'Universit\`{a} and Sezione INFN, Padova, Italy\\
$^{28}$ Dipartimento di Fisica `E.R.~Caianiello' dell'Universit\`{a} and Gruppo Collegato INFN, Salerno, Italy\\
$^{29}$ Dipartimento DISAT del Politecnico and Sezione INFN, Turin, Italy\\
$^{30}$ Dipartimento di Scienze MIFT, Universit\`{a} di Messina, Messina, Italy\\
$^{31}$ Dipartimento Interateneo di Fisica `M.~Merlin' and Sezione INFN, Bari, Italy\\
$^{32}$ European Organization for Nuclear Research (CERN), Geneva, Switzerland\\
$^{33}$ Faculty of Electrical Engineering, Mechanical Engineering and Naval Architecture, University of Split, Split, Croatia\\
$^{34}$ Faculty of Nuclear Sciences and Physical Engineering, Czech Technical University in Prague, Prague, Czech Republic\\
$^{35}$ Faculty of Physics, Sofia University, Sofia, Bulgaria\\
$^{36}$ Faculty of Science, P.J.~\v{S}af\'{a}rik University, Ko\v{s}ice, Slovak Republic\\
$^{37}$ Faculty of Technology, Environmental and Social Sciences, Bergen, Norway\\
$^{38}$ Frankfurt Institute for Advanced Studies, Johann Wolfgang Goethe-Universit\"{a}t Frankfurt, Frankfurt, Germany\\
$^{39}$ Fudan University, Shanghai, China\\
$^{40}$ Gangneung-Wonju National University, Gangneung, Republic of Korea\\
$^{41}$ Gauhati University, Department of Physics, Guwahati, India\\
$^{42}$ Helmholtz-Institut f\"{u}r Strahlen- und Kernphysik, Rheinische Friedrich-Wilhelms-Universit\"{a}t Bonn, Bonn, Germany\\
$^{43}$ Helsinki Institute of Physics (HIP), Helsinki, Finland\\
$^{44}$ High Energy Physics Group,  Universidad Aut\'{o}noma de Puebla, Puebla, Mexico\\
$^{45}$ Horia Hulubei National Institute of Physics and Nuclear Engineering, Bucharest, Romania\\
$^{46}$ HUN-REN Wigner Research Centre for Physics, Budapest, Hungary\\
$^{47}$ Indian Institute of Technology Bombay (IIT), Mumbai, India\\
$^{48}$ Indian Institute of Technology Indore, Indore, India\\
$^{49}$ INFN, Laboratori Nazionali di Frascati, Frascati, Italy\\
$^{50}$ INFN, Sezione di Bari, Bari, Italy\\
$^{51}$ INFN, Sezione di Bologna, Bologna, Italy\\
$^{52}$ INFN, Sezione di Cagliari, Cagliari, Italy\\
$^{53}$ INFN, Sezione di Catania, Catania, Italy\\
$^{54}$ INFN, Sezione di Padova, Padova, Italy\\
$^{55}$ INFN, Sezione di Pavia, Pavia, Italy\\
$^{56}$ INFN, Sezione di Torino, Turin, Italy\\
$^{57}$ INFN, Sezione di Trieste, Trieste, Italy\\
$^{58}$ Inha University, Incheon, Republic of Korea\\
$^{59}$ Institute for Gravitational and Subatomic Physics (GRASP), Utrecht University/Nikhef, Utrecht, Netherlands\\
$^{60}$ Institute of Experimental Physics, Slovak Academy of Sciences, Ko\v{s}ice, Slovak Republic\\
$^{61}$ Institute of Physics, Homi Bhabha National Institute, Bhubaneswar, India\\
$^{62}$ Institute of Physics of the Czech Academy of Sciences, Prague, Czech Republic\\
$^{63}$ Institute of Space Science (ISS), Bucharest, Romania\\
$^{64}$ Institut f\"{u}r Kernphysik, Johann Wolfgang Goethe-Universit\"{a}t Frankfurt, Frankfurt, Germany\\
$^{65}$ Instituto de Ciencias Nucleares, Universidad Nacional Aut\'{o}noma de M\'{e}xico, Mexico City, Mexico\\
$^{66}$ Instituto de F\'{i}sica, Universidade Federal do Rio Grande do Sul (UFRGS), Porto Alegre, Brazil\\
$^{67}$ Instituto de F\'{\i}sica, Universidad Nacional Aut\'{o}noma de M\'{e}xico, Mexico City, Mexico\\
$^{68}$ iThemba LABS, National Research Foundation, Somerset West, South Africa\\
$^{69}$ Jeonbuk National University, Jeonju, Republic of Korea\\
$^{70}$ Korea Institute of Science and Technology Information, Daejeon, Republic of Korea\\
$^{71}$ Laboratoire de Physique Subatomique et de Cosmologie, Universit\'{e} Grenoble-Alpes, CNRS-IN2P3, Grenoble, France\\
$^{72}$ Lawrence Berkeley National Laboratory, Berkeley, California, United States\\
$^{73}$ Lund University Department of Physics, Division of Particle Physics, Lund, Sweden\\
$^{74}$ Marietta Blau Institute, Vienna, Austria\\
$^{75}$ Nagasaki Institute of Applied Science, Nagasaki, Japan\\
$^{76}$ Nara Women{'}s University (NWU), Nara, Japan\\
$^{77}$ National and Kapodistrian University of Athens, School of Science, Department of Physics , Athens, Greece\\
$^{78}$ National Centre for Nuclear Research, Warsaw, Poland\\
$^{79}$ National Institute of Science Education and Research, Homi Bhabha National Institute, Jatni, India\\
$^{80}$ National Nuclear Research Center, Baku, Azerbaijan\\
$^{81}$ National Research and Innovation Agency - BRIN, Jakarta, Indonesia\\
$^{82}$ Niels Bohr Institute, University of Copenhagen, Copenhagen, Denmark\\
$^{83}$ Nikhef, National institute for subatomic physics, Amsterdam, Netherlands\\
$^{84}$ Nuclear Physics Group, STFC Daresbury Laboratory, Daresbury, United Kingdom\\
$^{85}$ Nuclear Physics Institute of the Czech Academy of Sciences, Husinec-\v{R}e\v{z}, Czech Republic\\
$^{86}$ Oak Ridge National Laboratory, Oak Ridge, Tennessee, United States\\
$^{87}$ Ohio State University, Columbus, Ohio, United States\\
$^{88}$ Physics department, Faculty of science, University of Zagreb, Zagreb, Croatia\\
$^{89}$ Physics Department, Panjab University, Chandigarh, India\\
$^{90}$ Physics Department, University of Jammu, Jammu, India\\
$^{91}$ Physics Program and International Institute for Sustainability with Knotted Chiral Meta Matter (WPI-SKCM$^{2}$), Hiroshima University, Hiroshima, Japan\\
$^{92}$ Physikalisches Institut, Eberhard-Karls-Universit\"{a}t T\"{u}bingen, T\"{u}bingen, Germany\\
$^{93}$ Physikalisches Institut, Ruprecht-Karls-Universit\"{a}t Heidelberg, Heidelberg, Germany\\
$^{94}$ Physik Department, Technische Universit\"{a}t M\"{u}nchen, Munich, Germany\\
$^{95}$ Politecnico di Bari and Sezione INFN, Bari, Italy\\
$^{96}$ Research Division and ExtreMe Matter Institute EMMI, GSI Helmholtzzentrum f\"ur Schwerionenforschung GmbH, Darmstadt, Germany\\
$^{97}$ Saga University, Saga, Japan\\
$^{98}$ Saha Institute of Nuclear Physics, Homi Bhabha National Institute, Kolkata, India\\
$^{99}$ School of Physics and Astronomy, University of Birmingham, Birmingham, United Kingdom\\
$^{100}$ Secci\'{o}n F\'{\i}sica, Departamento de Ciencias, Pontificia Universidad Cat\'{o}lica del Per\'{u}, Lima, Peru\\
$^{101}$ SUBATECH, IMT Atlantique, Nantes Universit\'{e}, CNRS-IN2P3, Nantes, France\\
$^{102}$ Sungkyunkwan University, Suwon City, Republic of Korea\\
$^{103}$ Suranaree University of Technology, Nakhon Ratchasima, Thailand\\
$^{104}$ Technical University of Ko\v{s}ice, Ko\v{s}ice, Slovak Republic\\
$^{105}$ The Henryk Niewodniczanski Institute of Nuclear Physics, Polish Academy of Sciences, Cracow, Poland\\
$^{106}$ The University of Texas at Austin, Austin, Texas, United States\\
$^{107}$ Universidad Aut\'{o}noma de Sinaloa, Culiac\'{a}n, Mexico\\
$^{108}$ Universidade de S\~{a}o Paulo (USP), S\~{a}o Paulo, Brazil\\
$^{109}$ Universidade Estadual de Campinas (UNICAMP), Campinas, Brazil\\
$^{110}$ Universidade Federal do ABC, Santo Andre, Brazil\\
$^{111}$ Universitatea Nationala de Stiinta si Tehnologie Politehnica Bucuresti, Bucharest, Romania\\
$^{112}$ University of Cape Town, Cape Town, South Africa\\
$^{113}$ University of Derby, Derby, United Kingdom\\
$^{114}$ University of Houston, Houston, Texas, United States\\
$^{115}$ University of Jyv\"{a}skyl\"{a}, Jyv\"{a}skyl\"{a}, Finland\\
$^{116}$ University of Kansas, Lawrence, Kansas, United States\\
$^{117}$ University of Liverpool, Liverpool, United Kingdom\\
$^{118}$ University of Science and Technology of China, Hefei, China\\
$^{119}$ University of Silesia in Katowice, Katowice, Poland\\
$^{120}$ University of South-Eastern Norway, Kongsberg, Norway\\
$^{121}$ University of Tennessee, Knoxville, Tennessee, United States\\
$^{122}$ University of the Witwatersrand, Johannesburg, South Africa\\
$^{123}$ University of Tokyo, Tokyo, Japan\\
$^{124}$ University of Tsukuba, Tsukuba, Japan\\
$^{125}$ Universit\"{a}t M\"{u}nster, Institut f\"{u}r Kernphysik, M\"{u}nster, Germany\\
$^{126}$ Universit\'{e} Clermont Auvergne, CNRS/IN2P3, LPC, Clermont-Ferrand, France\\
$^{127}$ Universit\'{e} de Lyon, CNRS/IN2P3, Institut de Physique des 2 Infinis de Lyon, Lyon, France\\
$^{128}$ Universit\'{e} de Strasbourg, CNRS, IPHC UMR 7178, F-67000 Strasbourg, France, Strasbourg, France\\
$^{129}$ Universit\'{e} Paris-Saclay, Centre d'Etudes de Saclay (CEA), IRFU, D\'{e}partment de Physique Nucl\'{e}aire (DPhN), Saclay, France\\
$^{130}$ Universit\'{e}  Paris-Saclay, CNRS/IN2P3, IJCLab, Orsay, France\\
$^{131}$ Universit\`{a} degli Studi di Foggia, Foggia, Italy\\
$^{132}$ Universit\`{a} del Piemonte Orientale, Vercelli, Italy\\
$^{133}$ Universit\`{a} di Brescia, Brescia, Italy\\
$^{134}$ Variable Energy Cyclotron Centre, Homi Bhabha National Institute, Kolkata, India\\
$^{135}$ Warsaw University of Technology, Warsaw, Poland\\
$^{136}$ Wayne State University, Detroit, Michigan, United States\\
$^{137}$ Yale University, New Haven, Connecticut, United States\\
$^{138}$ Yildiz Technical University, Istanbul, Turkey\\
$^{139}$ Yonsei University, Seoul, Republic of Korea\\
$^{140}$ Affiliated with an institute formerly covered by a cooperation agreement with CERN\\
$^{141}$ Affiliated with an international laboratory covered by a cooperation agreement with CERN.\\

\end{flushleft}